\documentclass[preprint,showpacs,amsmath,amssymb,prd,nofootinbib]{revtex4}
\usepackage{epsf}
\usepackage{dcolumn}
\usepackage{amsmath}
\usepackage{amssymb}
\usepackage{graphicx}

\usepackage[breaklinks=true,colorlinks=true]{hyperref}

\begin{document}
\title{On the 256-dimensional gamma matrix representation of the Clifford algebra $\textit{C}\ell^{\texttt{R}}$(1,7) and its relation to the Lie algebra SO(1,9)}

\author{Volodimir Simulik $^{1,2}$, Ilona Vyikon $^{1}$}

\email{vsimulik@gmail.com}

\affiliation{$^{1}$  Institute of Electron Physics, National Academy of Sciences of Ukraine, 21 Universitetska Str. 88017 Uzhgorod, Ukraine}
\affiliation{$^{2}$ Erwin Shr\"odinger International Institute for Mathematics and Physics, University of Vienna}
\date{May 2022}
\begin{abstract}
\small
Extended gamma matrix Clifford--Dirac and SO(1,9) algebras in the terms of $8 \times 8$ matrices have been considered. The 256-dimensional gamma matrix representation of Clifford algebra for 8-component Dirac equation is suggested. Two isomorphic realizations $\textit{C}\ell^{\texttt{R}}$(0,8) and $\textit{C}\ell^{\texttt{R}}$(1,7) are considered. The corresponding gamma matrix representations of 45-dimensional SO(10) and SO(1,9) algebras, which contain standard and additional spin operators, are introduced as well. The SO(10), SO(1,9) and the corresponding $\textit{C}\ell^{\texttt{R}}$(0,8)$, \textit{C}\ell^{\texttt{R}}$(1,7) representations are determined as algebras over the field of real numbers. The suggested gamma matrix representations of the Lie algebras SO(10), SO(1,9) are constructed on the basis of the Clifford algebras $\textit{C}\ell^{\texttt{R}}$(0,8)$, \textit{C}\ell^{\texttt{R}}$(1,7) representations. Comparison with the corresponded algebras in the space of standard 4-component Dirac spinors is demonstrated. The proposed mathematical objects allow generalization of our results, obtained earlier for the standard Dirac equation, for equations of higher spin and, especially, for equations, describing particles with spin 3/2. The maximal 84-dimensional pure matrix algebra of invariance of the 8-component Dirac equation in the Foldy--Wouthuysen representation is found. The corresponding symmetry of the Dirac equation in ordinary representation is found as well. The possible generalizations of considered Lie algebras to the arbitrary dimensional SO(n) and SO(m,n) are discussed briefly.  
\normalsize
\end{abstract}
\maketitle

\section{Introduction}

The development of our investigations [1--4], new results and our future goals are presented. We consider the 8-component Dirac equation in the corresponding rigged Hilbert space, where the solutions of such Dirac equation are defined. The 8-component Dirac equation in the terms of $8\times 8$ gamma matrices has its own important and independent role in modern quantum field theory, which enhances our interest to the corresponding representations $\textit{C}\ell^{\texttt{R}}$(0,8)$, \textit{C}\ell^{\texttt{R}}$(1,7) of Clifford and SO(10), SO(1,9) algebras. The algebras found here have much higher dimensions than the previously found in [1--4] similar algebras for the 4-component Dirac equation. The suggested gamma matrix representations of the Lie algebras SO(10), SO(1,9) are constructed on the basis of the Clifford algebras $\textit{C}\ell^{\texttt{R}}$(0,8)$, \textit{C}\ell^{\texttt{R}}$(1,7) representations.

The 256-dimensional gamma matrix algebras (constructed with the help of additional operators) are put into consideration. Such matrix algebras are isomorphic to the  Clifford algebras $\textit{C}\ell^{\texttt{R}}$(0,8) and $\textit{C}\ell^{\texttt{R}}$(1,7) over the field of real numbers. Related representations of SO(10) and SO(1,9) algebras over the field of real numbers are considered as well. The role of mathematical objects proposed here for the generalization of our results, obtained earlier for the standard four-component Dirac equation, for the equations of higher spin and, especially, for equations, describing particles with spin 3/2, is considered.

At first similar algebras for the standard 4-component Dirac equation have been considered. The 64-dimensional gamma matrix representation of the corresponding Clifford algebra over the field of real numbers and the related representation of the SO(8) Lie algebra were introduced in Refs. [1--4] (the review and the final description are given in Ref. [5]). $\textit{C}\ell^{\texttt{R}}$(4,2) and $\textit{C}\ell^{\texttt{R}}$(0,6) realizations were considered. The role of matrix representations of such algebras in the quantum field theory was also investigated in Refs. [1--5]. An example of the standard Dirac equation was considered. Mathematical basis for our algebraic considerations was taken from Refs. [6--10]. Therefore, in Refs. [1--5] and here below we continued our 25 years study, in which different representations of the Clifford--Dirac algebra, useful in mathematical physics (see, e.g., [11, 12]), were introduced.

Unfortunatelly, in our papers [1--4] on this subject some non-correct interpretation of the Clifford algebras was given. Therefore, some reconsideration here can be useful.

Our publications [1--4] are not only about representations of the Clifford and SO(8) algebras. On this basis we suggested the method of derivation of the Bose (not Fermi) symmetries, solutions and conservation laws for the Dirac equation with nonzero mass. We started in Ref. [1] from the Lorentz and Poincar\'e symmetries. In [2, 3] we considered different realizations of the representations of the Lorentz and the Poincar\'e groups, with respect to which the Dirac equation ($m \neq 0$) is invariant. Ref. [4] is devoted to relationships between the representations of the Clifford and SO(8) algebras. On this basis in Refs. [13, 14] we were able to find new (including Bose) symmetries of the relativistic hydrogen atom, i.e. for the Dirac equation in an external Coulomb field. It is the essence of our Fermi-Bose dualism conception for the Dirac equation [1--3, 5]. 

Moreover, a 31-dimensional Lie algebra SO(6)$\oplus i\gamma^{0}$SO(6)$\oplus i\gamma^{0}$, which is the maximal pure matrix algebra of invariance of the Dirac equation interacting with a Coulomb field in the Foldy--Wouthuysen representation, was found in Refs. [13, 14]. The corresponding symmetry of the ordinary Dirac equation for the relativistic hydrogen atom was found as well, but the main part of the symmetry operators are not pure matrix.

Note that in addition to these main results Refs. [1--4] deal with the comparison of the Clifford and SO(8) algebras gamma matrix representations. We started from the study of Refs. [15, 16], where 16 elements of the Clifford-Dirac algebra are linked to 15 elements of the SO(3,3) Lie algebra. In our papers [1--4] we were able to show a relationship between the gamma matrix representation of the $\textit{C}\ell^{\texttt{R}}$(1,3) algebra and 15 elements of the SO(1,5) Lie algebra. On this basis, relationship between the representations of the $\textit{C}\ell^{\texttt{R}}$(0,6) and SO(8) argebras was considered [1--4]. Here below relationships $\textit{C}\ell^{\texttt{R}}$(0,8)$\Leftrightarrow$SO(10) and $\textit{C}\ell^{\texttt{R}}$(1,7)$\Leftrightarrow$SO(1,9) are investigated.

The proof of the Fermi--Bose duality of the 4-component Dirac equation for the case $m=0$ in the paper [17] was much easier. We used the ordinary 16-dimensional Clifford--Dirac algebra, the  Pauli--G$\ddot{\mathrm{u}}$rsey--Ibragimov transformations [18--20] and did not appeal to the Foldy--Wouthuysen representation as it occurred later in the general case of nonzero mass [1--4].  

For the 8-component Dirac equation, contrary to Refs. [1--4], much more extended representations of the Clifford and SO(10), SO(1,9) algebras can be found. Below we consider (constructed with the help of additional operators) the gamma matrix algebras, which are isomorphic to the Clifford algebras $\textit{C}\ell^{\texttt{R}}$(0,8) and $\textit{C}\ell^{\texttt{R}}$(1,7) over the field of real numbers. The corresponding representations of SO(10) and SO(1,9) algebras are considered as well. Relationships between such representations of the Clifford and SO(10), SO(1,9) algebras are investigated. Such new representations of these algebras over the field of real numbers will be useful for additional studies of the first-order 8-component partial differential equations of the quantum field theory, see, e.g., [21, 22], especially for the equation without redundant components for the spin 3/2 particle, suggested recently in Refs. [5, 23, 24].

Note that here we not consider generalizations of SO(10), SO(1,9) for an arbitrary SO(n), SO(m,n) algebras representations. We fixed only gamma matrix representations of the Lie algebras SO(10) and SO(1,9), which are necessary for our further investigations of the equations from [5, 23, 24].

The finally mentioned suggestion for the possibility of SO(n), SO(m,n) is only the indication of the way.  

\section{Notations, assumptions and definitions}

One of the principal objects of the relativistic quantum mechanics and field theory is the Dirac equation, see, e.g. [25]. This equation describes a particle-antiparticle doublet with spins s=(1/2,1/2) or, in other words, spin 1/2 fermion-antifermion doublet. Presentation of the Dirac equation operator $D\equiv i\gamma^{\mu}\partial_{\mu}-m$, as well as other operators of quantum spinor field, in terms of $\gamma$ matrices allows one to use the anti-commutation relations between the Clifford algebra operators directly for finding symmetries, solutions, conservation laws, performing the canonical quantization and calculation of the interaction processes in the quantum field models. An important fact is that application of the Clifford algebra essentially simplifies the calculations.

Thus, for the case of a free non-interacting spinor field the Dirac equation has the form
\begin{equation}
\label{eq1}
(i\gamma^{\mu}\partial_{\mu}-m)\psi(x)=0,
\end{equation}
where 
\begin{equation}
\label{eq2}
x\in \mathrm{M}(1,3), \quad \partial_{\mu}\equiv \partial/\partial x^{\mu}, \quad \mu=\overline{0,3}, \quad j=1,2,3,
\end{equation}
$\mathrm{M}(1,3)=\{x\equiv(x^{\mu})=(x^{0}=t, \, \overrightarrow{x}\equiv(x^{j}))\}$ is the Minkowski space-time and 4-component function $\psi(x)$ belongs to rigged Hilbert space
\begin{equation}
\label{eq3}
\mathrm{S}^{3,4}\subset\mathrm{H}^{3,4}\subset\mathrm{S}^{3,4*}.
\end{equation}
Here the first index denotes the dimension of the space of arguments, the second index denotes the dimension of the space of functions. Note that due to the special role of the time variable $x^{0}=t\in (x^{\mu})$  (in obvious analogy with nonrelativistic theory), in general consideration one can use the quantum-mechanical rigged Hilbert space (3). Here the Schwartz test function space $\mathrm{S}^{3,4}$ is dense in the Schwartz generalized function space $\mathrm{S}^{3,4*}$  and $\mathrm{H}^{3,4}$ is the quantum-mechanical Hilbert space of 4-component functions over $\mathrm{R}^{3}\subset \mathrm{M}(1,3)$. For the manifestly covariant field theory the rigged Hilbert space is taken as $\mathrm{S}^{4,4}\subset\mathrm{H}^{4,4}\subset\mathrm{S}^{4,4*}$.

Starting from Sec.3 we choose the 8-component spinors and the rigged Hilbert space $\mathrm{S}^{3,8}\subset\mathrm{H}^{3,8}\subset\mathrm{S}^{3,8*}$. 

In order to finish with notations, assumptions and definitions we note that here the system of units  $\varepsilon=\mu=\hbar=c=1$ is chosen, the metric tensor in Minkowski space-time $\mathrm{M}(1,3)$ is given by
\begin{equation}
\label{eq4}
g^{\mu\nu}=g_{\mu\nu}=g^{\mu}_{\nu}, \, \left(g^{\mu}_{\nu}\right)=\mathrm{diag}\left(1,-1,-1,-1\right); \quad x_{\mu}=g_{\mu\nu}x^{\mu},
\end{equation}
and summation over the twice repeated indices is implied.

The Dirac matrices $\gamma$ are taken in the standard Dirac--Pauli representation
\begin{equation}
\label{eq5} \gamma^{0}=\left| {{\begin{array}{*{20}c}
 \mathrm{I}_{2} \hfill &  0 \hfill\\
 0 \hfill & -\mathrm{I}_{2}  \hfill\\
 \end{array} }} \right|, \quad \gamma^{\ell}=\left| {{\begin{array}{*{20}c}
 0 \hfill &  \sigma^{\ell} \hfill\\
 -\sigma^{\ell} \hfill & 0  \hfill\\
 \end{array} }} \right|, \quad \ell=1,2,3, 
\end{equation}
where the Pauli matrices are given by
\begin{equation}
\label{eq6}
\sigma^{1}=\left| {{\begin{array}{*{20}c}
 0 \hfill &  1 \hfill\\
 1 \hfill & 0  \hfill\\
 \end{array} }} \right|, \quad \sigma^{2}=\left| {{\begin{array}{*{20}c}
 0 \hfill &  -i \hfill\\
 i \hfill &   0  \hfill\\
 \end{array} }} \right|, \quad \sigma^{3}=\left| {{\begin{array}{*{20}c}
 1 \hfill &  0 \hfill\\
 0 \hfill & -1  \hfill\\
\end{array} }} \right|,
\end{equation}
$\sigma^{1}\sigma^{2}=i\sigma^{3}, \, \sigma^{2}\sigma^{3}=i\sigma^{1}, \, \sigma^{3}\sigma^{1}=i\sigma^{2}$. Four matrix operators (5) satisfy the anti-commutation relations of the Clifford algebra
\begin{equation}
\label{eq7} \gamma^{\mu}\gamma^{\nu}+\gamma^{\nu}\gamma^{\mu}=2g^{\mu\nu}, \quad \gamma_{\mu}=g_{\mu\nu}\gamma^{\nu}, \,  \gamma^{\dagger\ell}=-\gamma^{\ell},  \,  \gamma^{\dagger 0}=\gamma^{0},
\end{equation}
and realize the matrix representation of the 16-dimensional ($2^{4}$=16) Clifford algebra $\textit{C}\ell^{\texttt{C}}$(1,3) over the field of complex numbers.

Indeed, the rule of recalculation of the basic elements can be found, e.g. in Ref. [26]. The corresponding set is as follows:
$$\gamma^0, \gamma^1, \gamma^2, \gamma^3, \quad \gamma^0\gamma^1, \gamma^0 \gamma^2,\gamma^0\gamma^3,\gamma^1 \gamma^2, \gamma^1 \gamma^3,\gamma^2\gamma^3,$$
\begin{equation}
\label{eq8}
\gamma^0\gamma^1\gamma^2,\gamma^0\gamma^1\gamma^3,\gamma^0\gamma^2\gamma^3,\gamma^1\gamma^2\gamma^3, \quad \gamma^0\gamma^1\gamma^2\gamma^3, \quad \mathrm{I}_{4}.
\end{equation}
We have a hope that similar recalculations for the considered below algebras in Appendices 1, 2 will be useful for physicists.

For our purposes we introduce an additional matrix
\begin{equation}
\label{eq9} \gamma^{4}=\gamma^{0}\gamma^{1}\gamma^{2}\gamma^{3}=-i\left| {{\begin{array}{*{20}c}
 0 \hfill & \mathrm{I}_{2} \hfill\\
 \mathrm{I}_{2} \hfill & 0  \hfill\\
 \end{array} }} \right|, \quad \mathrm{I}_{2}=\left| {{\begin{array}{*{20}c}
 1 \hfill &  0 \hfill\\
 0 \hfill & 1  \hfill\\
\end{array} }} \right|.
\end{equation}
The anti-commutation relations of the Clifford algebra includes this matrix operator
\begin{equation}
\label{eq10} \gamma^{\bar{\mu}}\gamma^{\bar{\nu}}+\gamma^{\bar{\nu}}\gamma^{\bar{\mu}}=2g^{\bar{\mu}\bar{\nu}}, \quad \bar{\mu},\bar{\nu}=\overline{0,4}, \, (g^{\bar{\mu}\bar{\nu}})=(+----),  
\end{equation}
and the metric tensor is given by $5\times 5$ matrix. Despite the fact that $\gamma^{4}=\gamma^{0}\gamma^{1}\gamma^{2}\gamma^{3}$ the operator $\gamma^{4}$ is linearly independent. Indeed, it is the independent member of the set (8).

Here and in our publications (see, e. g. the Refs. [1--5]) we use the anti-Hermitian matrix $\gamma^{4}=\gamma^{0}\gamma^{1}\gamma^{2}\gamma^{3}$ instead of the Hermitian matrix $\gamma^{5}_{\mathrm{standard}}$ of other authors [25]. Our $\gamma^{4}$ is equal to $i\gamma^{5}_{\mathrm{standard}}$. Notation $\gamma^{5}$ is used in Refs. [1--5] and below for a completely different matrix operator $\gamma^{5}\equiv \gamma^{1}\gamma^{3}\hat{C}$.

We use the definition of the Clifford algebra from Refs. [6, 7].

\textbf{Remark 1.} Below in Sec.4 and Sec.5 we put into consideration new matrix algebras in the terms of $8\times 8$ gamma matrices. It is necessary to give the proof that the dimensions of such algebras are equal to $2^{8}=256$ and this objects are isomorphic to the $\textit{C}\ell^{\texttt{R}}$(0,8) and $\textit{C}\ell^{\texttt{R}}$(1,7) Clifford algebras. We have a hope that the place of such proof should be chosen here on the basis of the well known example. The proof in Sec.4 and Sec.5 is similar.

\textit{Briefly.} At first, it is necessary to take into account formulae (7)--(10) and the evident fact that the spur of all matrices (5), (9) is equal to zero. The next step is to prove that all matrices in the set $\left\{\gamma^{0}, \gamma^{1}, \gamma^{2}, \gamma^{3}, \gamma^{4}\right\}$ are linearly independent. Indeed, the equation $\sum x_{\bar{\mu}}\gamma^{\bar{\mu}}=0$ yields if and only if $x_{0}=x_{1}=x_{2}=x_{3}=x_{4}$. Of course, the same assertion is true for the set $\left\{\gamma^{0}, \gamma^{1}, \gamma^{2}, \gamma^{3}\right\}$. It may seems that the dimension of this gamma matrix algebra is equal to $2^{5}=32$! The solution follows after taking into account the equality $\gamma^{4}=\gamma^{0}\gamma^{1}\gamma^{2}\gamma^{3}$. Indeed, $\gamma^{0}\gamma^{4}= \gamma^{1}\gamma^{2}\gamma^{3}, \, \gamma^{0}\gamma^{1}\gamma^{4}= \gamma^{2}\gamma^{3}, \, \gamma^{0}\gamma^{1}\gamma^{2}\gamma^{4}= -\gamma^{3}$, which are already inside the set (8).

Therefore, the dimension of gamma matrix algebra under consideration is $2^{4}=16$. This algebra according to [6] is isomorphic to geometric algebra $\textit{C}\ell^{\texttt{C}}$(1,3).

\section{Representations of the $\textit{C}\ell^{\texttt{R}}$(0,6) and SO(8) algebras for the ordinary Dirac equation}

Consider a set of seven $\gamma$ matrices
\begin{equation}
\label{eq11}
\gamma^{1},\,\gamma^{2},\,\gamma^{3},\,\gamma^{4}=\gamma^{0}\gamma^{1}\gamma^{2}\gamma^{3},\,\gamma^{5}=\gamma^{1}\gamma^{3}\hat{C},
\,\gamma^{6}=i\gamma^{1}\gamma^{3}\hat{C},\,\gamma^{7}=i\gamma^{0},
\end{equation}
where $ \gamma ^0, \, \gamma ^{\ell}$ are given in formulae (5), $\sigma ^k$ are the standard Pauli matrices (6), imaginary unit $i=\sqrt{-1}$ is considered as an operator, $\hat{C}$ is the operator of complex conjugation, $\hat{C}\psi=\psi^{*}$ (the operator of involution in $\mathrm{H}^{3,4}$), i.e. here these operators are nontrivial elements of the algebra. The conjugation operatr $\hat{C}$ is conjugate linear. It means that $\hat{C}i =-i\hat{C}$ (see articles [1--4] for details) and, therefore, for example, $i\gamma^{5}=-\gamma^{5}i$. In any case we do not touch here the complex field algebra among the results, as well as in the papers [1--4]. All suggested algebras are over the field of real numbers.

Operators (11) satisfy the anti-commutation relations 
\begin{equation}
\label{eq12} \gamma ^{\hat{\mathrm{A}}} \gamma ^{\hat{\mathrm{B}}} + \gamma
^{\hat{\mathrm{B}}}\gamma ^{\hat{\mathrm{A}}} = -
2\delta^{\hat{\mathrm{A}}\hat{\mathrm{B}}},\quad \hat{\mathrm{A}},\hat{\mathrm{B}}=\overline{1,7},
\end{equation}
for the generators of representation of the Clifford algebra over the field of real numbers. Furthermore, for the matrices (11) the following relation is valid: $\gamma^{1}\gamma^{2}\gamma^{3}\gamma^{4}\gamma^{5}\gamma^{6}\gamma^{7}=\mathrm{I}_{4}$. 

Over the real field all seven generators (3.1) are linearly independent. Nevertheless, in complete analogy with Remark 1 and relation $\gamma^{4}=\gamma^{0}\gamma^{1}\gamma^{2}\gamma^{3}$ one must take into account the relation $\gamma^{7}=i\gamma^{0}=-i\gamma^{1}\gamma^{2}\gamma^{3}\gamma^{4}$. Therefore. we deal here with the 64-dimensional gamma matrix algebra, which is isomorphic to the $\textit{C}\ell^{\texttt{R}}$(0,6) over the field of real numbers ($2^{6}=64$).

The first 16 operators of such algebra are given in (5) and (8), the next 16 ones are found from them by multiplication by imaginary unit $i=\sqrt{-1}$. The rest 32 operators are found from first 32 ones by multiplication by operator $\hat{C}$ of complex conjugation. Thus, if the notation "stand CD" ("stand" and CD are taken from standard Clifford--Dirac) is introduced for the set of 16 matrices (5) and (8), then the set of 64 elements of $\textit{C}\ell^{\texttt{R}}$(0,6) algebra representation will be given by
\begin{equation}
\label{eq13}
\left\{ ({\mbox{stand \, CD}})\cup i\cdot({\mbox{stand \, CD}})\cup \hat{C}\cdot({\mbox{stand \, CD}})\cup i\hat{C}\cdot({\mbox{stand \, CD}}) \right\}.
\end{equation}
We have a hope that another way of recalculation of these 64 elements, which is given in Appendix 1, will be useful for physicists.

In Refs. [4, 5] we consider another 64-dimensional realization $\textit{C}\ell^{\texttt{R}}$(4,2) as well.

Operators (11) also generate 28 matrices 
\begin{equation}
\label{eq14}
s^{\widetilde{\mathrm{A}}\widetilde{\mathrm{B}}}=\{s^{\hat{\mathrm{A}}\hat{\mathrm{B}}}=\frac{1}{4}[\gamma
^{\hat{\mathrm{A}}},\gamma
^{\hat{\mathrm{B}}}],\,s^{\hat{\mathrm{A}}8}=-s^{8\hat{\mathrm{A}}}=\frac{1}{2}\gamma
^{\hat{\mathrm{A}}}\},
\end{equation}
$\widetilde{\mathrm{A}},\widetilde{\mathrm{B}}=\overline{1,8}, \, \hat{\mathrm{A}},\hat{\mathrm{B}}=\overline{1,7},$ which satisfy the commutation relations of the Lie algebra SO(8)
\begin{equation}
\label{eq15}
[s^{\widetilde{\mathrm{A}}\widetilde{\mathrm{B}}},s^{\widetilde{\mathrm{C}}\widetilde{\mathrm{D}}}]=
\delta^{\widetilde{\mathrm{A}}\widetilde{\mathrm{C}}}s^{\widetilde{\mathrm{B}}\widetilde{\mathrm{D}}}
+\delta^{\widetilde{\mathrm{C}}\widetilde{\mathrm{B}}}s^{\widetilde{\mathrm{D}}\widetilde{\mathrm{A}}}
+\delta^{\widetilde{\mathrm{B}}\widetilde{\mathrm{D}}}s^{\widetilde{\mathrm{A}}\widetilde{\mathrm{C}}}
+\delta^{\widetilde{\mathrm{D}}\widetilde{\mathrm{A}}}s^{\widetilde{\mathrm{C}}\widetilde{\mathrm{B}}}.
\end{equation}
It is evident that here we also have an algebra over the field of real numbers. Note that here (as in Refs. [1--4]) the anti-Hermitian realization of the SO(10) operators is chosen, for the reasons see, e.g., Refs. [1--5] and [9, 10]. We appeal to the anti-Hermitian realizations of the generators starting from Refs. [11, 12].

The explicit form of the 28 elements of the $\gamma$ matrix representation of the SO(8) algebra is given in Table 1.
$$\mathrm{\textbf{Table 1.} \, The \, 28 \, \, elements \, of \, the \, \gamma \, \, matrix \,\, representation \, of \, the \, \, SO(8) \, algebra}$$
\begin{center}
\begin{tabular}{|c|c|c|c|c|c|c|}
\hline
\rule{0pt}{5mm} $\frac{1}{2}\gamma^{1}\gamma^{2}$  & $\frac{1}{2}\gamma^{1}\gamma^{3}$ & $\frac{1}{2}\gamma^{1}\gamma^{4}$ & $\frac{1}{2}\gamma^{1}\gamma^{5}$ & $\frac{1}{2}\gamma^{1}\gamma^{6}$ & $\frac{1}{2}\gamma^{1}\gamma^{7}$ & $s^{18}\equiv \frac{1}{2}\gamma^{1}$ \\
\hline
\rule{0pt}{5mm} & $\frac{1}{2}\gamma^{2}\gamma^{3}$ & $\frac{1}{2}\gamma^{2}\gamma^{4}$  & $\frac{1}{2}\gamma^{2}\gamma^{5}$ & $\frac{1}{2}\gamma^{2}\gamma^{6}$ & $\frac{1}{2}\gamma^{2}\gamma^{7}$   & $s^{28}\equiv \frac{1}{2}\gamma^{2}$ \\
\hline
\rule{0pt}{5mm} &  & $\frac{1}{2}\gamma^{3}\gamma^{4}$  & $\frac{1}{2}\gamma^{3}\gamma^{5}$ & $\frac{1}{2}\gamma^{3}\gamma^{6}$ & $\frac{1}{2}\gamma^{3}\gamma^{7}$ & $s^{38}\equiv \frac{1}{2}\gamma^{3}$ \\
\hline
\rule{0pt}{5mm}  &  &  & $\frac{1}{2}\gamma^{4}\gamma^{5}$ & $\frac{1}{2}\gamma^{4}\gamma^{6}$ & $\frac{1}{2}\gamma^{4}\gamma^{7}$ & $s^{48}\equiv \frac{1}{2}\gamma^{4}$\\
\hline
\rule{0pt}{5mm} &  &   &  & $\frac{1}{2}\gamma^{5}\gamma^{6}$ & $\frac{1}{2}\gamma^{5}\gamma^{7}$ &  $s^{58}\equiv \frac{1}{2}\gamma^{5}$ \\
\hline
\rule{0pt}{5mm} &  &  &  &  & $\frac{1}{2}\gamma^{6}\gamma^{7}$ &  $s^{68}\equiv \frac{1}{2}\gamma^{6}$ \\
\hline
\rule{0pt}{5mm} &  &  &  &  &  &  $s^{78}\equiv \frac{1}{2}\gamma^{7}$ \\
\hline
\end{tabular}
\label{Table_1}
\end{center}

\vspace{0,1in}

The first application of the gamma matrix representations of the algebras $\textit{C}\ell^{\texttt{R}}$(0,6) and SO(8) is symmetry analysis (search for groups and algebras with respect to which the equation is invariant) of the Dirac equation with nonzero mass. It is easy to understand that the Foldy--Wouthuysen representation [27, 28] is preferable for such analysis. Indeed, in this representation one should calculate the commutation relations of possible pure matrix symmetry operators from the set (14) and Table 1 only with the matrix part $i\gamma^{0}$ of the Foldy--Wouthuysen equation operator: $(\partial _0 +i\gamma^{0}\sqrt{-\Delta + m^{2}} -\frac{e^{2}}{\left|\vec{x}\right|})\phi (x) = 0$. Having the symmetries of the Foldy--Wouthuysen equation one can find the symmetries of the Dirac equation on the basis of the inverse Foldy--Wouthuysen transformation. Note that after such transformation only a small part of the symmetry operators will be pure matrix, the main part of the operators will contain a nonlocal operator $\omega \equiv \sqrt{-\Delta + m^{2}}$ and functions of it. A 31-dimensional Lie algebra SO(6)$\oplus i\gamma^{0}$SO(6)$\oplus i\gamma^{0}$ was found, which is formed by the elements from $\textit{C}\ell^{\texttt{R}}$(0,6) and is the maximal pure matrix algebra of invariance of the Dirac equation in the Foldy--Wouthuysen representation, see, e.g., Refs. [1--4, 13, 14]. The nonlocal image of SO(6)$\oplus i\gamma^{0}$SO(6)$\oplus i\gamma^{0}$ algebra in the Dirac representation is given in Refs. [13, 14].

Furthermore, in the Foldy--Wouthuysen representation two different subsets of operators ($s^{23}, \, s^{31}, \, s^{12}$) and ($s^{56},\,s^{64},\,s^{45}$) from Table 1 (i) determine two different sets of SU(2) spin 1/2 generators, (ii) commute between each other and (iii) commute with the operator of the Dirac equation in the Foldy--Wouthuysen representation. On this bases in Refs. [1--4] Bose symmetries, Bose solutions and Bose conservation laws for the Dirac equation with nonzero mass (1) were found. Among the Bose-symmetries the important Lorentz and Poincar$\mathrm{\acute{e}}$ algebras of invariance of the Dirac equation with nonzero mass were found.

A transition to the 8-component Dirac equation and $8 \times 8$ gamma matrices opens wider possibilities and can be used for the equations of higher spins.

\section{Representation of the $\textit{C}\ell^{\texttt{R}}$(0,8) algebra for the 8-component Dirac equation}

Consider a set of nine $8\times 8$ $\Gamma$ matrices
$$\Gamma^{j}
 = \left|
\begin{array}{cccc}
0 & 0 & 0 & \sigma^{j}\\
0 & 0 & \sigma^{j} & 0\\
0 & \sigma^{j} & 0 & 0\\
\sigma^{j} & 0 & 0 & 0\\
\end{array} \right|, \, j=1,2,3, \quad \Gamma^{4}=i\left|
\begin{array}{cccc}
0 & 0 & 0 & -\mathrm{I}_{2}\\
0 & 0 & \mathrm{I}_{2} & 0\\
0 & -\mathrm{I}_{2} & 0 & 0\\
\mathrm{I}_{2} & 0 & 0 & 0\\
\end{array} \right|, $$
$$\Gamma^{5}=i\left|
\begin{array}{cccc}
0 & 0 & -\mathrm{I}_{2} & 0\\
0 & 0 & 0 & -\mathrm{I}_{2}\\
\mathrm{I}_{2} & 0 & 0 & 0\\
0 & \mathrm{I}_{2} & 0 & 0\\
\end{array} \right|,$$
\begin{equation}
\label{eq16}
\Gamma^{6}=\left|
\begin{array}{cccc}
0 & 0 & \mathrm{I}_{2} & 0\\
0 & 0 & 0 & -\mathrm{I}_{2}\\
\mathrm{I}_{2} & 0 & 0 & 0\\
0 & -\mathrm{I}_{2} & 0 & 0\\
\end{array} \right|, \quad \Gamma^{7}=\left|
\begin{array}{cccc}
\mathrm{I}_{2} & 0 & 0 & 0\\
0 & \mathrm{I}_{2} & 0 & 0\\
0 & 0 & -\mathrm{I}_{2} & 0\\
0 & 0 & 0 & -\mathrm{I}_{2}\\
\end{array} \right|, \quad \mathrm{I}_{2}=\left| {{\begin{array}{*{20}c}
 1 \hfill &  0 \hfill\\
 0 \hfill & 1  \hfill\\
\end{array} }} \right|,
\end{equation}
$$\Gamma^{8}=\left|
\begin{array}{cccc}
0 & 0 & -\sigma^{2}\widehat{C} & 0\\
0 & 0 & 0 & \sigma^{2}\widehat{C}\\
\sigma^{2}\widehat{C} & 0 & 0 & 0\\
0 & -\sigma^{2}\widehat{C} & 0 & 0\\
\end{array} \right|, \, \Gamma^{9}=\left|
\begin{array}{cccc}
0 & 0 & -i\sigma^{2}\widehat{C} & 0\\
0 & 0 & 0 & i\sigma^{2}\widehat{C}\\
i\sigma^{2}\widehat{C} & 0 & 0 & 0\\
0 & -i\sigma^{2}\widehat{C} & 0 & 0\\
\end{array} \right|,$$
where $\sigma^{j}$ are the standard Pauli matrices (6) and $\widehat{C}$ is the operator of complex conjugation, $\widehat{C}\psi=\psi^{*}$ (the operator of involution in the Hilbert space $\mathrm{H}^{3,2}$). The operators (16) satisfy the anti-commutation relations 
\begin{equation}
\label{eq17}
\Gamma ^\mathrm{A} \Gamma ^\mathrm{B} + \Gamma
^\mathrm{B}\Gamma ^\mathrm{A} = 2\delta^{\mathrm{A}\mathrm{B}},\quad \mathrm{A},\mathrm{B}=\overline{1,9},
\end{equation}
of the Clifford algebra. It is useful to take into account the relationships
\begin{equation}
\label{eq18}
\Gamma^{1}\Gamma^{2}\Gamma^{3}\Gamma^{4}\Gamma^{5}\Gamma^{6}\Gamma^{7}\Gamma^{8}\Gamma^{9}=\mathrm{I}_{8}, \quad \Gamma^{7}=-i\Gamma^{1}\Gamma^{2}\Gamma^{3}\Gamma^{4}\Gamma^{5}\Gamma^{6}.
\end{equation}
Here nine operators from the set (16) are linearly independent. In complete analogy with Remark 1 and relation $\gamma^{4}=\gamma^{0}\gamma^{1}\gamma^{2}\gamma^{3}$ one must take into account the second relation from (18) and determine the generators and the basis of the corresponding gamma matrix algebra. This algebra is isomorphic to the Clifford algebra $\textit{C}\ell^{\texttt{R}}$(8,0) over the field of real numbers. Corresponding dimension is $2^{8}=256$. 

Moreover, we again can take into account that $\Gamma^{7}=-i\Gamma^{1}\Gamma^{2}\Gamma^{3}\Gamma^{4}\Gamma^{5}\Gamma^{6}$ and consider the subalgebra of (16), which is determined by first six generators from the set (16). This gamma matrix algebra is isomorphic to 64-dimensional Clifford algebra $\textit{C}\ell^{\texttt{C}}$(6,0) over the field of complex numbers (here we deal with $2^{6}=64$ dimensions).  Note that such algebra in this formalism is an analogue of the ordinary 16-dimensional Clifford--Dirac algebra $\textit{C}\ell^{\texttt{C}}$(1,3) (see, e.g. Sec. 2).

Therefore, similarly to Refs. [1--5], we can recalculate the complete set of 256 elements of the gamma matrix algebra, which is the subalgebra of algebra (16), and which is isomorphic to the 256-dimensional Clifford algebra $\textit{C}\ell^{\texttt{R}}$(0,8) over the field of real numbers, in the form
\begin{equation}
\label{eq19} \left\{ (64\textit{C}\ell^{\texttt{C}}(0,6))\cup i\cdot(64\textit{C}\ell^{\texttt{C}}(0,6))\cup \widetilde{C}\cdot(64\textit{C}\ell^{\texttt{C}}(0,6))\cup i\widetilde{C}\cdot(64\textit{C}\ell^{\texttt{C}}(0,6)) \right\},
\end{equation}
where $\widetilde{C}$ is the operator of complex conjugation in the space of solutions of 8-component Dirac equation, $\widetilde{C}\psi=\psi^{*}$ (the operator of involution in the Hilbert space $\mathrm{H}^{3,8}$).

Thus, we use the following idea here. We work not only with matrices but with matrices and operators acting on matrices, and use these operators together with the matrices to generate the higher dimensional gamma matrix algebras being isomorphic to Clifford algebras. An example of how fruitful this idea might be was provided not only in [1], but in [29] as well.

It is useful to present the basis (19) visually in an explicit form, as one can find, e.g., in Ref. [26] for the standard representation of $\textit{C}\ell^{\texttt{C}}$(0,4) of the Clifford--Dirac algebra (see the formulas (8) here for $\textit{C}\ell^{\texttt{C}}$(1,3)) representation. Despite the fact that standard Grassmann basises for $\textit{C}\ell^{\texttt{R}}$(0,6) and $\textit{C}\ell^{\texttt{R}}$(0,8) algebras are well known, we describe in Appendices 1 and 2 the corresponding basises of isomorphic to them $\Gamma$-matrix algebras. The analogy with (8) from Ref. [26] is used. Moreover, we are sure that such Appendices will be useful for physicists. 

\section{Representation of the $\textit{C}\ell^{\texttt{R}}$(1,7) algebra for the 8-component Dirac equation}

Let us mark the first difference between the consideration in the four-component formalism of [1--4] and the eight-component formalism considered here. In the space of 4-component Dirac $\psi$-function it was impossible to introduce the representation $\textit{C}\ell^{\texttt{R}}$(1,5) instead (or together with) $\textit{C}\ell^{\texttt{R}}$(0,6). Here such option is possible and is under consideration due to the additional generators of the algebra $\textit{C}\ell^{\texttt{R}}$(1,7):
$$\quad \Gamma^{0}=\left|
\begin{array}{cccc}
\mathrm{I}_{2} & 0 & 0 & 0\\
0 & \mathrm{I}_{2} & 0 & 0\\
0 & 0 & -\mathrm{I}_{2} & 0\\
0 & 0 & 0 & -\mathrm{I}_{2}\\
\end{array} \right|, \quad \Gamma^{j}
 = \left|
\begin{array}{cccc}
0 & 0 & 0 & \sigma^{j}\\
0 & 0 & \sigma^{j} & 0\\
0 & -\sigma^{j} & 0 & 0\\
-\sigma^{j} & 0 & 0 & 0\\
\end{array} \right|,$$
$$\Gamma^{4}=i\left|
\begin{array}{cccc}
0 & 0 & 0 & -\mathrm{I}_{2}\\
0 & 0 & \mathrm{I}_{2} & 0\\
0 & \mathrm{I}_{2} & 0 & 0\\
-\mathrm{I}_{2} & 0 & 0 & 0\\
\end{array} \right|,$$
\begin{equation}
\label{eq20}
\Gamma^{5}=i\left|
\begin{array}{cccc}
0 & 0 & -\mathrm{I}_{2} & 0\\
0 & 0 & 0 & -\mathrm{I}_{2}\\
-\mathrm{I}_{2} & 0 & 0 & 0\\
0 & -\mathrm{I}_{2} & 0 & 0\\
\end{array} \right|, \quad \Gamma^{6}=\left|
\begin{array}{cccc}
0 & 0 & \mathrm{I}_{2} & 0\\
0 & 0 & 0 & -\mathrm{I}_{2}\\
-\mathrm{I}_{2} & 0 & 0 & 0\\
0 & \mathrm{I}_{2} & 0 & 0\\
\end{array} \right|,
\end{equation}
$$\Gamma^{7}=\left|
\begin{array}{cccc}
0 & 0 & \sigma^{2}\widehat{C} & 0\\
0 & 0 & 0 & -\sigma^{2}\widehat{C}\\
\sigma^{2}\widehat{C} & 0 & 0 & 0\\
0 & -\sigma^{2}\widehat{C} & 0 & 0\\
\end{array} \right|, \, \Gamma^{8}=\left|
\begin{array}{cccc}
0 & 0 & i\sigma^{2}\widehat{C} & 0\\
0 & 0 & 0 & -i\sigma^{2}\widehat{C}\\
i\sigma^{2}\widehat{C} & 0 & 0 & 0\\
0 & -i\sigma^{2}\widehat{C} & 0 & 0\\
\end{array} \right|.$$
The operators (20) satisfy the anti-commutation relations 
\begin{equation}
\label{eq21}
\Gamma ^\mathrm{\tilde{A}} \Gamma ^\mathrm{\tilde{B}} + \Gamma
^\mathrm{\tilde{B}}\Gamma ^\mathrm{\tilde{A}} = 2g^{\mathrm{\tilde{A}}\mathrm{\tilde{B}}}, \quad g=(+--------), \quad \mathrm{\tilde{A}},\mathrm{\tilde{B}}=\overline{0,8},
\end{equation}
of the Clifford algebra generators. Again, according to Remark 1, we must choose eight elements from the set (20) of linearly independent operators. The analogy with the previous section is valid. Therefore, this 256-dimensional gamma matrix algebra is isomorphic to the Clifford algebra $\textit{C}\ell^{\texttt{R}}$(1,7) over the field of real numbers.

Note that according to [6] (page 217) the algebras considered in Sec.4 and here in Sec.5 are isomorphic. Indeed, $\textit{C}\ell^{\texttt{R}}$(0,8)$\cong$$\textit{C}\ell^{\texttt{R}}$(1,7)$\cong$Mat(16,$\texttt{R}$). Nevertheless, both matrix agebras (16) and (20) can be useful, as we can conclude from the experience [1--4], in the proof of Fermi--Bose duality of the 8-component Dirac equation. 

Further description of this algebra is similar to the one given above in Sec 4.

Using the algebra (20), (21) as an example, let's pay attention that gamma matrices (20) have the form of B14, page 443, in [30]. Moreover, these matrices can be presented in block form in the terms of $4 \times 4$ Pauli matrices. Nevertheless, the splitting as B16 from [30] is not necessary for the purposes of this article explained in the Abstract and Introduction. Below we are applying the objects like $8 \times 8$ matrices and this dimension is necessary. Our final goal is to construct three independent sets of SU(2) generators (see the text above the equation (24)).  

\section{Gamma matrix representation of SO(10) algebra}

Operators (16) also generate 45 matrices 
\begin{equation}
\label{eq22}
s^{\breve{\mathrm{A}}\breve{\mathrm{B}}}=\{s^{\mathrm{A}\mathrm{B}}=-\frac{1}{4}[\Gamma
^\mathrm{A},\Gamma
^\mathrm{B}],\,s^{\mathrm{A}10}=-s^{10\mathrm{A}}=-\frac{1}{2}\Gamma
^\mathrm{A}\},
\end{equation}
$\breve{\mathrm{A}},\breve{\mathrm{B}}=\overline{1,10}, \, \mathrm{A},\mathrm{B}=\overline{1,9}$, which satisfy the commutation relations of the generators of the Lie algebra SO(10):
\begin{equation}
\label{eq23}
[s^{\breve{\mathrm{A}}\breve{\mathrm{B}}},s^{\breve{\mathrm{C}}\breve{\mathrm{D}}}]=
\delta^{\breve{\mathrm{A}}\breve{\mathrm{C}}}s^{\breve{\mathrm{B}}\breve{\mathrm{D}}}
+\delta^{\breve{\mathrm{C}}\breve{\mathrm{B}}}s^{\breve{\mathrm{D}}\breve{\mathrm{A}}}
+\delta^{\breve{\mathrm{B}}\breve{\mathrm{D}}}s^{\breve{\mathrm{A}}\breve{\mathrm{C}}}
+\delta^{\breve{\mathrm{D}}\breve{\mathrm{A}}}s^{\breve{\mathrm{C}}\breve{\mathrm{B}}}.
\end{equation}
Note that here as in Sec.3 the anti-Hermitian realization of the SO(10) operators is chosen, for the reasons see, e.g., [1--5, 9, 10].

The explicit form of the 45 elements of the $\Gamma$ matrix representation of the SO(10) algebra is given in Table 2. \textit{Note that all elements of this table should be multiplied by the factor (-1/2). The example is in bottom corner.}
$$\mathrm{\textbf{Table 2.} \, The \, 45 \, \, elements \, of \, the \, \Gamma \, \, matrix \,\, representation \, of \, the \, \, SO(10) \, algebra}$$
\small
\begin{center}
\begin{tabular}{|c|c|c|c|c|c|c|c|c|}
\hline
\rule{0pt}{1mm} $-\frac{1}{2}\Gamma^{1}\Gamma^{2}$  & $-\frac{1}{2}\Gamma^{1}\Gamma^{3}$ & $-\frac{1}{2}\Gamma^{1}\Gamma^{4}$ & $-\frac{1}{2}\Gamma^{1}\Gamma^{5}$ & $-\frac{1}{2}\Gamma^{1}\Gamma^{6}$ & $-\frac{1}{2}\Gamma^{1}\Gamma^{7}$ & $-\frac{1}{2}\Gamma^{1}\Gamma^{8}$ & $-\frac{1}{2}\Gamma^{1}\Gamma^{9}$ & $s^{1 10}\equiv -\frac{1}{2}\Gamma^{1}$ \\
\hline
\rule{0pt}{1mm} & $-\frac{1}{2}\Gamma^{2}\Gamma^{3}$ & $-\frac{1}{2}\Gamma^{2}\Gamma^{4}$  & $-\frac{1}{2}\Gamma^{2}\Gamma^{5}$ & $-\frac{1}{2}\Gamma^{2}\Gamma^{6}$ & $-\frac{1}{2}\Gamma^{2}\Gamma^{7}$   & $-\frac{1}{2}\Gamma^{2}\Gamma^{8}$ & $-\frac{1}{2}\Gamma^{2}\Gamma^{9}$ & $s^{2 10}\equiv -\frac{1}{2}\Gamma^{2}$ \\
\hline
\rule{0pt}{1mm} &  & $-\frac{1}{2}\Gamma^{3}\Gamma^{4}$  & $-\frac{1}{2}\Gamma^{3}\Gamma^{5}$ & $-\frac{1}{2}\Gamma^{3}\Gamma^{6}$ & $-\frac{1}{2}\Gamma^{3}\Gamma^{7}$ & $-\frac{1}{2}\Gamma^{3}\Gamma^{8}$ & $-\frac{1}{2}\Gamma^{3}\Gamma^{9}$ & $s^{3 10}\equiv -\frac{1}{2}\Gamma^{3}$  \\
\hline
\rule{0pt}{1mm}  &  &  & $-\frac{1}{2}\Gamma^{4}\Gamma^{5}$ & $-\frac{1}{2}\Gamma^{4}\Gamma^{6}$ & $-\frac{1}{2}\Gamma^{4}\Gamma^{7}$ & $-\frac{1}{2}\Gamma^{4}\Gamma^{8}$ & $-\frac{1}{2}\Gamma^{4}\Gamma^{9}$ & $s^{4 10}\equiv -\frac{1}{2}\Gamma^{4}$ \\
\hline
\rule{0pt}{1mm} &  &   &  & $-\frac{1}{2}\Gamma^{5}\Gamma^{6}$ & $-\frac{1}{2}\Gamma^{5}\Gamma^{7}$ & $-\frac{1}{2}\Gamma^{5}\Gamma^{8}$ & $-\frac{1}{2}\Gamma^{5}\Gamma^{9}$ & $s^{5 10}\equiv -\frac{1}{2}\Gamma^{5}$ \\
\hline
\rule{0pt}{1mm} &  &  &  &  & $-\frac{1}{2}\Gamma^{6}\Gamma^{7}$ & $-\frac{1}{2}\Gamma^{6}\Gamma^{8}$ & $-\frac{1}{2}\Gamma^{6}\Gamma^{9}$ & $s^{6 10}\equiv -\frac{1}{2}\Gamma^{6}$ \\
\hline
\rule{0pt}{1mm} &  &  &  &  &  &  $-\frac{1}{2}\Gamma^{7}\Gamma^{8}$ &  $-\frac{1}{2}\Gamma^{7}\Gamma^{9}$ & $s^{7 10}\equiv -\frac{1}{2}\Gamma^{7}$ \\
\hline
\rule{0pt}{1mm} &  &  &  &  &  &  &  $-\frac{1}{2}\Gamma^{8}\Gamma^{9}$ & $s^{8 10}\equiv -\frac{1}{2}\Gamma^{8}$ \\
\hline
\rule{0pt}{1mm} &  &  &  &  &  &  &  &  $s^{9 10}\equiv -\frac{1}{2}\Gamma^{9}$ \\
\hline
\end{tabular}
\label{Table_2}
\end{center}
\normalsize

The gamma matrices in Table 2 are taken from the set (16).

The dimension of the corresponding SO(n) algebra is given by $\frac{n(n-1)}{2}=\frac{10\cdot 9}{2}=45$. Therefore, here we deal with SO(10) algebra representation.

Table 2 demonstrates not only the explicit form of the generators (22) but also the commutation relations (23). Indeed, it is evident that the generators with different indices commute between each other. Further, it is evident that here we have three independent sets of SU(2) generators ($s^{1}=s^{23},\,s^{2}=s^{31},\,s^{3}=s^{12}$), which commute between each other. They are given by the following operators from Table 2: ($-\frac{1}{2}\Gamma^{2}\Gamma^{3},-\frac{1}{2}\Gamma^{3}\Gamma^{1},-\frac{1}{2}\Gamma^{1}\Gamma^{2}$), ($-\frac{1}{2}\Gamma^{5}\Gamma^{6},-\frac{1}{2}\Gamma^{6}\Gamma^{4},-\frac{1}{2}\Gamma^{4}\Gamma^{5}$), ($-\frac{1}{2}\Gamma^{8}\Gamma^{9},-\frac{1}{2}\Gamma^{9}\Gamma^{7},-\frac{1}{2}\Gamma^{7}\Gamma^{8}$). All above mention sets of SU(2) generators commute with the operator of the 8-component Foldy--Wouthuysen equation in anti-Hermitian form
\begin{equation}
\label{eq24}
(\partial _0 +i\Gamma^{0}\sqrt{-\Delta + m^{2}} -\frac{e^{2}}{\left|\vec{x}\right|})\phi (x) = 0; \quad  x\in \mathrm{M}(1,3), \; \phi\in \left\{\mathrm{S}^{3,8}\subset\mathrm{H}^{3,8}\subset\mathrm{S}^{3,8*}\right\}.
\end{equation}
Note that in Refs. [1--5] in the representation of SO(8) for 4-component spinors we have only two independent SU(2) sets, combinations of which unable us to prove the Bose symmetries of the Dirac equation. Here, similarly, due to the presence of the triplet of SU(2) sets the spin 3/2 Lorentz and Poincar\'e symmetries for the equation suggested in [5, 23, 24] can be found. Of course, the Bose symmetries of the 8-component Dirac equation can be found as well.

The above considered Lie algebra of the SO(10) group can be generalised to the SO(n) algebra of an arbotrary dimension:
\begin{equation}
\label{eq25}
s^{\breve{p}\breve{q}}=\{s^{pq}=-\frac{1}{4}[\Gamma
^{p},\Gamma
^{q}],\,s^{p \, q+1}=-s^{q+1 \, p}=-\frac{1}{2}\Gamma
^{p}\},
\end{equation}
where $\breve{p},\breve{q}=\overline{1,n+1}, \, p,q=\overline{1,n}$. This is different from similar constructions in [31]. Indeed, we use here the anti-Hermitian infinitesimal operators (25). Moreover, we include in the set (25) the gamma matrices such like $\gamma^{4}=\gamma^{0}\gamma^{1}\gamma^{2}\gamma^{3}$ and $\Gamma^{7}=-i\Gamma^{1}\Gamma^{2}\Gamma^{3}\Gamma^{4}\Gamma^{5}\Gamma^{6}$ from (18). The main difference from [31] consists in the inclusion in (25) of gamma matrices, which contain additional operator $\widehat{C}$ (see, e.g., matrices $\Gamma^{8}$ and $\Gamma^{9}$ from the set (16)). 

\section{Gamma matrix representation of SO(1,9) algebra}

The explicit form of the corresponding generators follows from the set (20). 45 gamma matrix generators of SO(1,9) algebra are given by  
\begin{equation}
\label{eq26}
s^{\widehat{\mathrm{A}}\widehat{\mathrm{B}}}=\{s^{\tilde{\mathrm{A}}\tilde{\mathrm{B}}}=\frac{1}{4}[\Gamma
^{\tilde{\mathrm{A}}},\Gamma
^{\tilde{\mathrm{B}}}],\,s^{\tilde{\mathrm{A}}9}=-s^{9\tilde{\mathrm{A}}}=\frac{1}{2}\Gamma
^{\tilde{\mathrm{A}}}\}, 
\end{equation}
where $\widehat{\mathrm{A}},\widehat{\mathrm{B}}=\overline{0,9}, \, \mathrm{\tilde{A}},\mathrm{\tilde{B}}=\overline{0,8}$. Matrix operators (26) satisfy the commutation relations of the generators of the Lie algebra SO(1,9)
\begin{equation}
\label{eq27}
[s^{\widehat{\mathrm{A}}\widehat{\mathrm{B}}},s^{\widehat{\mathrm{C}}\widehat{\mathrm{D}}}]=
-g^{\widehat{\mathrm{A}}\widehat{\mathrm{C}}}s^{\widehat{\mathrm{B}}\widehat{\mathrm{D}}}
-g^{\widehat{\mathrm{C}}\widehat{\mathrm{B}}}s^{\widehat{\mathrm{D}}\widehat{\mathrm{A}}}
-g^{\widehat{\mathrm{B}}\widehat{\mathrm{D}}}s^{\widehat{\mathrm{A}}\widehat{\mathrm{C}}}
-g^{\widehat{\mathrm{D}}\widehat{\mathrm{A}}}s^{\widehat{\mathrm{C}}\widehat{\mathrm{B}}},
\end{equation}
where the metric tensor $g$ is already given in relations (21).

Note that here similarly to Sec. 4 for the same reasons the anti-Hermitian realization of the SO(1,9) operators is chosen.

The explicit form of the 45 elements of the $\Gamma$ matrix representation of the SO(1,9) algebra is given in Table 3.
$$\mathrm{\textbf{Table 3.} \, 45 \, \, elements \, of \, the \, \Gamma \, \, matrix \,\, representation \, of \, the \, \, SO(1,9) \, algebra}$$
\small
\begin{center}
\begin{tabular}{|c|c|c|c|c|c|c|c|c|c|}
\hline
\rule{0pt}{5mm} $\frac{1}{2}\Gamma^{0}\Gamma^{1}$  & $\frac{1}{2}\Gamma^{0}\Gamma^{2}$ & $\frac{1}{2}\Gamma^{0}\Gamma^{3}$ & $\frac{1}{2}\Gamma^{0}\Gamma^{4}$ & $\frac{1}{2}\Gamma^{0}\Gamma^{5}$ & $\frac{1}{2}\Gamma^{0}\Gamma^{6}$ & $\frac{1}{2}\Gamma^{0}\Gamma^{7}$ & $\frac{1}{2}\Gamma^{0}\Gamma^{8}$ & $s^{0 9}\equiv \frac{1}{2}\Gamma^{0}$ \\
\hline
\rule{0pt}{5mm} & $\frac{1}{2}\Gamma^{1}\Gamma^{2}$ & $\frac{1}{2}\Gamma^{1}\Gamma^{3}$  & $\frac{1}{2}\Gamma^{1}\Gamma^{4}$ & $\frac{1}{2}\Gamma^{1}\Gamma^{5}$ & $\frac{1}{2}\Gamma^{1}\Gamma^{6}$   & $\frac{1}{2}\Gamma^{1}\Gamma^{7}$ & $\frac{1}{2}\Gamma^{1}\Gamma^{8}$ & $s^{1 9}\equiv \frac{1}{2}\Gamma^{1}$ \\
\hline
\rule{0pt}{5mm} &  & $\frac{1}{2}\Gamma^{2}\Gamma^{3}$  & $\frac{1}{2}\Gamma^{2}\Gamma^{4}$ & $\frac{1}{2}\Gamma^{2}\Gamma^{5}$ & $\frac{1}{2}\Gamma^{2}\Gamma^{6}$ & $\frac{1}{2}\Gamma^{2}\Gamma^{7}$ & $\frac{1}{2}\Gamma^{2}\Gamma^{8}$ & $s^{2 9}\equiv \frac{1}{2}\Gamma^{2}$  \\
\hline
\rule{0pt}{5mm}  &  &  & $\frac{1}{2}\Gamma^{3}\Gamma^{4}$ & $\frac{1}{2}\Gamma^{3}\Gamma^{5}$ & $\frac{1}{2}\Gamma^{3}\Gamma^{6}$ & $\frac{1}{2}\Gamma^{3}\Gamma^{7}$ & $\frac{1}{2}\Gamma^{3}\Gamma^{8}$ & $s^{3 9}\equiv \frac{1}{2}\Gamma^{3}$ \\
\hline
\rule{0pt}{5mm} &  &   &  & $\frac{1}{2}\Gamma^{4}\Gamma^{5}$ & $\frac{1}{2}\Gamma^{4}\Gamma^{6}$ & $\frac{1}{2}\Gamma^{4}\Gamma^{7}$ & $\frac{1}{2}\Gamma^{4}\Gamma^{8}$ & $s^{4 9}\equiv \frac{1}{2}\Gamma^{4}$ \\
\hline
\rule{0pt}{5mm} &  &  &  &  & $\frac{1}{2}\Gamma^{5}\Gamma^{6}$ & $\frac{1}{2}\Gamma^{5}\Gamma^{7}$ & $\frac{1}{2}\Gamma^{5}\Gamma^{8}$ & $s^{5 9}\equiv \frac{1}{2}\Gamma^{5}$ \\
\hline
\rule{0pt}{5mm} &  &  &  &  &  &  $\frac{1}{2}\Gamma^{6}\Gamma^{7}$ &  $\frac{1}{2}\Gamma^{6}\Gamma^{8}$ & $s^{6 9}\equiv \frac{1}{2}\Gamma^{6}$ \\
\hline
\rule{0pt}{5mm} &  &  &  &  &  &  &  $\frac{1}{2}\Gamma^{7}\Gamma^{8}$ & $s^{7 9}\equiv \frac{1}{2}\Gamma^{7}$ \\
\hline
\rule{0pt}{5mm} &  &  &  &  &  &  &  &  $s^{8 9}\equiv \frac{1}{2}\Gamma^{8}$ \\
\hline
\end{tabular}
\label{Table_3}
\end{center}
\normalsize

\vspace{0,1in}

The gamma matrices in Table 3 are taken from the set (20).

The dimension of the corresponding SO(m,n) algebra is given by the following formula $\frac{(m+n)(m+n-1)}{2}=\frac{10\cdot 9}{2}=45$. Therefore, here we deal with the representation of SO(1,9) algebra.

Table 3 demonstrates both the explicit form of the generators (26) and the commutation relations (27). Indeed, it is evident that generators with different indices commute between each other. Here again we have three independent sets of SU(2) generators ($s^{1}=s^{23},\,s^{2}=s^{31},\,s^{3}=s^{12}$), which commute between each other. Therefore, here similarly to the consideration in Sec. 6 the spin 3/2 Lorentz and Poincar\'e symmetries for the equation suggested in [5, 23, 24] can be found.

Similarly to the formulae in (25), above considered Lie algebra of the SO(1,9) group can be generalised to the SO(n, m) algebra of an arbotrary dimension:
\begin{equation}
\label{eq28}
s^{\breve{p}\breve{q}}=\{s^{pq}=\frac{1}{4}[\Gamma
^{p},\Gamma
^{q}],\,s^{p \, q+1}=-s^{q+1 \, p}=-\frac{1}{2}\Gamma
^{p}\},
\end{equation}
where $\breve{p},\breve{q}=\overline{0,n+1}, \, p,q=\overline{0,n}$.

\section{Application to symmetry analysis}

Consider the 21-dimensional representation of the subalgebra SO(7) of the algebra SO(10):
\begin{equation}
\{s^{\hat{\mathrm{A}}\hat{\mathrm{B}}}\}=\{s^{\hat{\mathrm{A}}\hat{\mathrm{B}}}\equiv\frac{1}{4}[\Gamma^{\hat{\mathrm{A}}},\Gamma^{\hat{\mathrm{B}}}]\}, \quad \hat{\mathrm{A}},\hat{\mathrm{B}}=\overline{1,7},
\label{eq29}
\end{equation}
Indeed, it is the pure matrix algebra of invariance of the Dirac equation in the Foldy--Wouthuysen representation (24).
The complete set of the pure matrix symmetries of equation (6.3) is given by 21 elements of SO(7) plus three SU(2) operators ($s^{1}_{\mathrm{III}}=-\frac{1}{2}\Gamma^{8}\Gamma^{9}, \, s^{2}_{\mathrm{III}}=-\frac{1}{2}\Gamma^{9}\Gamma^{7}, \, s^{3}_{\mathrm{III}}=-\frac{1}{2}\Gamma^{7}\Gamma^{8}$). The corresponding symmetries of the Dirac equation can be found by inverse Foldy--Wouthuysen transformation [27] in the space of 8-component spinors, see, e.g. [5]. Note that in the Dirac representation the main part of these operators will not be pure matrix. The usefulness of the Foldy--Wouthuysen transformation is demonstrated recently in Refs. [32, 33] as well as in our above mentioned papers. 

The 84-dimesinal set of pure matrix operators can be found by multiplication of 21 elements of SO(7) by each element from the given here set SU(2). However, it will be an overlapping algebra. Note that sometimes an overlapping algebra can be useful as well. We can recall a 31-dimensinal algebra C(1,3)$\oplus\varepsilon$C(1,3)$\oplus\varepsilon$, where $\varepsilon$ is the duality transformation (the transformation of Heaviside--Larmor--Rainich [34--36] in the space of field-strengths of electromagnetic field). Such maximal first-order symmetry of the Maxwell equations in the terms of field strengths was found in Ref. [37]. The usefulness was demonstrated e.g. in Ref. [38] and in many papers by other authors, which, unfortunately, often forgott to refer on the result from [37].

\section{Brief conclusions}

The above presented consideration is an introduction to the proof of the Fermi--Bose duality property of the 8-component Dirac equation with nonzero mass. Note that in the corresponding space of 8-component solutions the concept of the Fermi--Bose dualism of the Dirac equation is richer than for the standard Dirac equation in Refs. [1--4]. Here such concept contains (i) spin 1/2 properties, (ii) spin 1 (Bose) properties and (iii) spin 3/2 properties.

Derivation of the Fermi-Bose duality property of the ordinary Dirac equation in Refs. [1--4] was based on assertion that in the Foldy--Wouthuysen representation two subsets  ($s^{23},\,s^{31},\,s^{12}$) and ($s^{56},\,s^{64},\,s^{45}$) of operators $s^{\widetilde{\mathrm{A}}\widetilde{\mathrm{B}}}$ from Table 1 (i) determine two different sets of SU(2) spin 1/2 generators, (ii) commute between each other and (iii) commute with the operator of the Dirac equation in the Foldy--Wouthuysen representation $(\partial _0 +i\gamma^{0}\sqrt{-\Delta + m^{2}} -\frac{e^{2}}{\left|\vec{x}\right|})\phi (x) = 0$. Indeed, on this basis in Refs. [1--4] the Bose (spin 1) symmetries, Bose solutions and Bose conservation laws for the Dirac equation with nonzero mass (1) were found. Among the Bose symmetries the important Lorentz and Poincar$\mathrm{\acute{e}}$ algebras of invariance of the Dirac equation with nonzero mass were found.

Here (see, e.g. the Table 2) for the Dirac equation in the terms of 8x8 gamma matrices, more exactly for this equation in the Foldy--Wouthuysen representation (6.3) (as for the start), three such subsets of SU(2) generators are found: ($-\frac{1}{2}\Gamma^{2}\Gamma^{3},-\frac{1}{2}\Gamma^{3}\Gamma^{1},-\frac{1}{2}\Gamma^{1}\Gamma^{2}$), ($-\frac{1}{2}\Gamma^{5}\Gamma^{6},-\frac{1}{2}\Gamma^{6}\Gamma^{4},-\frac{1}{2}\Gamma^{4}\Gamma^{5}$), ($-\frac{1}{2}\Gamma^{8}\Gamma^{9},-\frac{1}{2}\Gamma^{9}\Gamma^{7},-\frac{1}{2}\Gamma^{7}\Gamma^{8}$). On this basis we can prove not only the spin 1 properties of the corresponding Dirac equation but the spin 3/2 properties as well. The method is known from Refs. [1--4]. Therefore, the suggested gamma matrix representations of different algebras open new possibilities for the investigations of the field theory equations for the higher spins, especially for the spin s=3/2. Note that well defined theory for the particles with spin s=3/2 is still under consideration [39--49] and is one of the problems of contemporary theoretical high energy physics.

Despite some generalizations of the representations SO(10), SO(1,9) to the forms (25), (28), the further development of the formalism to arbitrary dimensions is not the purpose of this article.

\section{Acknowledgements}

Volodimir Simulik is very grateful for the two month Fellowship at the Erwin Sch\"odinger International Institute for Mathematics and Physics of the University of Vienna. Authors are very grateful for unknown referees of the journal version of this paper.

\section{Appendix 1}

The set of 64 elements of the gamma matrix algebra, which is isomorphic to the Clifford algebra $\textit{C}\ell^{\texttt{R}}$(0,6) over the field of real numbers, is presented below. We have first six elements from the set (16), 15 elements as pairs of operators

\vspace{0,1in}

\begin{tabular}{ccccccccc}

\rule{0pt}{5mm} $\Gamma^{1}\Gamma^{2}$  & $\Gamma^{1}\Gamma^{3}$ & $\Gamma^{1}\Gamma^{4}$ & $\Gamma^{1}\Gamma^{5}$ & $\Gamma^{1}\Gamma^{6}$ \\

\rule{0pt}{5mm} & $\Gamma^{2}\Gamma^{3}$ & $\Gamma^{2}\Gamma^{4}$  & $\Gamma^{2}\Gamma^{5}$ & $\Gamma^{2}\Gamma^{6}$ \\

\rule{0pt}{5mm} &  & $\Gamma^{3}\Gamma^{4}$  & $\Gamma^{3}\Gamma^{5}$ & $\Gamma^{3}\Gamma^{6}$ \\

\rule{0pt}{5mm}  &  &  & $\Gamma^{4}\Gamma^{5}$ & $\Gamma^{4}\Gamma^{6}$ \\

\rule{0pt}{5mm} &  &   &  & $\Gamma^{5}\Gamma^{6}$, \\

\end{tabular}

\vspace{0,5in}

20 elements as operator triplets

\vspace{0,1in}

\begin{tabular}{cccccccccccccccc}

\rule{0pt}{5mm} $\Gamma^{1}\Gamma^{2}\Gamma^{3}$  & $\Gamma^{1}\Gamma^{2}\Gamma^{4}$ & $\Gamma^{1}\Gamma^{2}\Gamma^{5}$ & $\Gamma^{1}\Gamma^{2}\Gamma^{6}$ &  &  &  & $\Gamma^{2}\Gamma^{3}\Gamma^{4}$ & $\Gamma^{2}\Gamma^{3}\Gamma^{5}$ & $\Gamma^{2}\Gamma^{3}\Gamma^{6}$  \\

\rule{0pt}{5mm} & $\Gamma^{1}\Gamma^{3}\Gamma^{4}$ & $\Gamma^{1}\Gamma^{3}\Gamma^{5}$  & $\Gamma^{1}\Gamma^{3}\Gamma^{6}$ &  &  &  &  & $\Gamma^{2}\Gamma^{4}\Gamma^{5}$ & $\Gamma^{2}\Gamma^{4}\Gamma^{6}$ \\

\rule{0pt}{5mm} &  & $\Gamma^{1}\Gamma^{4}\Gamma^{5}$  & $\Gamma^{1}\Gamma^{4}\Gamma^{6}$ &  &  &  &  &  & $\Gamma^{2}\Gamma^{5}\Gamma^{6}$ \\

\rule{0pt}{5mm}  &  &  & $\Gamma^{1}\Gamma^{5}\Gamma^{6}$ \\

\rule{0pt}{5mm} & $\Gamma^{3}\Gamma^{4}\Gamma^{5}$ & $\Gamma^{3}\Gamma^{4}\Gamma^{6}$  &  &  &  &  & $\Gamma^{4}\Gamma^{5}\Gamma^{6}$ \\

\rule{0pt}{5mm} &  & $\Gamma^{3}\Gamma^{5}\Gamma^{6}$ \\

\end{tabular}

\vspace{0,1in}

15 elements as products of four operators

\vspace{0,1in}
\small
\begin{tabular}{ccccccccccccccccc}

\rule{0pt}{5mm} $\Gamma^{1}\Gamma^{2}\Gamma^{3}\Gamma^{4}$  & $\Gamma^{1}\Gamma^{2}\Gamma^{3}\Gamma^{5}$ & $\Gamma^{1}\Gamma^{2}\Gamma^{3}\Gamma^{6}$ & &  &  & $\Gamma^{1}\Gamma^{3}\Gamma^{4}\Gamma^{5}$ & $\Gamma^{1}\Gamma^{3}\Gamma^{4}\Gamma^{6}$ &  &  &  & $\Gamma^{1}\Gamma^{4}\Gamma^{5}\Gamma^{6}$ \\

\rule{0pt}{5mm} & $\Gamma^{1}\Gamma^{2}\Gamma^{4}\Gamma^{5}$ & $\Gamma^{1}\Gamma^{2}\Gamma^{4}\Gamma^{6}$  &  &  &  &  & $\Gamma^{1}\Gamma^{3}\Gamma^{5}\Gamma^{6}$ \\

\rule{0pt}{5mm} &  & $\Gamma^{1}\Gamma^{2}\Gamma^{5}\Gamma^{6}$ \\

\end{tabular}

\normalsize

\begin{tabular}{ccccccccccccccccc}

\rule{0pt}{5mm} $\Gamma^{2}\Gamma^{3}\Gamma^{4}\Gamma^{5}$  & $\Gamma^{2}\Gamma^{3}\Gamma^{4}\Gamma^{6}$  &  &  &  & $\Gamma^{2}\Gamma^{4}\Gamma^{5}\Gamma^{6}$  &  &  &  & $\Gamma^{3}\Gamma^{4}\Gamma^{5}\Gamma^{6}$ \\

\rule{0pt}{5mm} & $\Gamma^{2}\Gamma^{3}\Gamma^{5}\Gamma^{6}$,  \\

\end{tabular}

\vspace{0,1in}

6 elements as products of five operators

$$\begin{array}{cccc}
& \Gamma^{1}\Gamma^{2}\Gamma^{3}\Gamma^{4}\Gamma^{5}\ & \Gamma^{1}\Gamma^{2}\Gamma^{3}\Gamma^{4}\Gamma^{6}\\\
&   & \Gamma^{1}\Gamma^{2}\Gamma^{3}\Gamma^{5}\Gamma^{6}\\\
\end{array}$$
$$\Gamma^{1}\Gamma^{2}\Gamma^{4}\Gamma^{5}\Gamma^{6} \quad \Gamma^{1}\Gamma^{3}\Gamma^{4}\Gamma^{5}\Gamma^{6} \quad \Gamma^{2}\Gamma^{3}\Gamma^{4}\Gamma^{5}\Gamma^{6},$$

\vspace{0,1in}

one product of six operators $\Gamma^{1}\Gamma^{2}\Gamma^{3}\Gamma^{4}\Gamma^{5}\Gamma^{6}$ and $\mathrm{I}_{8}$ as the unit element of the algebra.

\section{Appendix 2}

The set of 256 elements of the gamma matrix algebra, which is isomorphic to the Clifford algebra $\textit{C}\ell^{\texttt{R}}$(0,8) over the field of real numbers,  can be recalculated similarly to the method presented in Appendix 1. We have eight independent elements from the set (4.1), 28 elements as pairs of operators, 56 elements as operator triplets, 70 products of four operators, 56 elements as products of five operators, 28 products of six operators, 8 products of seven operators, one product of eight operators $\Gamma^{1}\Gamma^{2}\Gamma^{3}\Gamma^{4}\Gamma^{5}\Gamma^{6}\Gamma^{7}\Gamma^{8}$ and $\mathrm{I}_{8}$ as the unit element of the algebra.




\end{document}